\documentstyle{article}
\begin{document}
\begin{center}{\bf Weinberg's energy-momentum pseudotensor for Schwarzschild
field} \end{center}
\begin{center} {\sl A.I.Nikishov}\end{center}
{\sl P.N.Lebedev Institute of Physics, Russian Academy of Science, 117924 Moscow,
Russia}
\begin{center} Summary\end{center}
Weinberg's energy-momentum pseudotensor is obtained for Schwarzschild metric in
harmonic coordinates. On the horizon it possesses unintegrable singularities.
For this reason the total energy of a collapsar can't be obtained by
integrating energy density over the system's volume. The implication for
gravity theories is noted. A thought on how to choose unique
energy-momentum tensor is given.

 \section{Introduction}

There is no energy-momentum tensor of gravitational field in general
relativity. Instead there are infinitely many pseudotensors [1,2]. The concept
of nonlocalizability of gravitational energy density was introduced to explain
this unusual situation. Yet this concept seems unnecessary, if the gravity
theory is build by field-theoretical means without requiring general covariance
or if one assumes the existence of privileged frame. In general relativity
the harmonic coordinates provide natural candidacy for privileged
coordinates for an isolated system [3].

 Weinberg's pseudotensor (the suffix pseudo is dropped in the following) is
singled out by the fact that it is the source of gravitational field [4].
For this reason it is interesting to find it in harmonic coordinates, which
goes over to Minkovskiian ones far away from gravitating body. Due to unusual
properties of space-time beyond horizon and interchanged role of time and
radial coordinates there [5], we expect that it will be impossible to get
the collapsar total energy by integrating energy density over system's
volume. Calculations confirm this: the energy-momentum tensor have
unintegrable singularities on the horizon.

\section{Calculation of energy-momentum tensor}

We use on the whole Weinberg notation, but denote harmonic coordinates
$x_i$, $\vert\vec x\vert=r$ with small letters. Indices of $h_{\mu\nu}$,
 $R^{(1)}_{\mu\nu}$,
 $\frac{\partial}{\partial
x^{\mu}}$ are raised and lowered with the help of  $\eta$, indices of
generally covariant tensors such as $R_{\mu\nu}$ are raised and
lowered with the help of $g$. Latin indices run from 1 to 3.
 $$
   g_{\mu\nu}=\eta_{\mu\nu}+h_{\mu\nu},\quad
  \eta_{\mu\nu}=diag(-1,1,1,1),\quad \phi=-\frac{GM}r, \eqno(1)
   $$
   $$
d\tau^2=-g_{\mu\nu}dx^{\mu}dx^{\nu}=\frac{1+\phi}{1-\phi}dt^2-(1-\phi)^2
d\vec x^2-\frac{1-\phi}{1+\phi}\phi^2\frac{(\vec xd\vec x)^2}{r^2}.\eqno(2)
$$
Now we reproduce eqs. (7.6.3) ³ (7.6.4) from [4]. The exact Einstein equations
are written there in the form
 $$
R^{(1)}_{\mu\kappa}-\frac12\eta_{\mu\kappa}R^{(1)}=-8\pi G[T_{\mu\kappa}+
t_{\mu\kappa}],\eqno(3)
$$
where
$$
t_{\mu\kappa}=\frac1{8\pi G}[R_{\mu\kappa}-\frac12g_{\mu\kappa}R-
R^{(1)}_{\mu\kappa}+\frac12\eta_{\mu\kappa}R^{(1)}],\eqno(4)
$$
and $R^{(1)}_{\mu\kappa}$ is linear in $h$ part of $R_{\mu\kappa}:$
 $$
R^{(1)}_{\mu\kappa}=\frac12[h_{,\mu\kappa}-h^{\lambda}{}_{\mu,\lambda\kappa}
-h^{\lambda}{}_{\kappa,\lambda\mu}+h_{\mu\kappa,\lambda}{}^{\lambda}],
\quad h_{,\mu}=\frac{\partial}{\partial x^{\mu}}h. \eqno(5)
$$
Eq.(3) has the form of wave equation for spin-2 field.
Its source is
$T_{\mu\kappa}+t_{\mu\kappa}$. Hence, $t_{\mu\kappa}$ (i.e. energy-momentum
tensor of gravitational field) is also a source of gravitational field.
Eq. (3) is suggested by solution of Einstein equations by iteration.
In linear approximation  $h_{\mu\nu}=h^{(1)_{\mu\nu}}$ is generated
by material tensor. Inserting this solution of linearized equation in (4) and
retaining quadratic in $h^{(1)}$ terms, we get
$t^{(2)}_{\mu\kappa}$.  [See eq.(7.6.14) in [4], in which it is not indicated
explicitly that figuring there $h_{\mu\nu}$ are $h^{(1)}_{\mu\nu}$.
The expression for $t^{(2)}_{\mu\nu}$, obtained from that eq., for Newtonian
center is given in [6].  It coincides with eqs. (11), (13) below, which were
found from exact expressions.] Further,
$t^{(2)}_{\mu\nu}$ according to wave equation generates $h^{(2)}_{\mu\nu}$
and so on. The sum over all approximations gives $h_{\mu\kappa}$, which exactly
satisfies Einstein equations.

 Now we assume up to eq.(14) that the matter is absent in the considered
region. Then the energy-momentum tensor has the form
 $$
 t_{\mu\nu}=\frac{1}{8\pi G}[\frac12\eta_{\mu\nu}R^{(1)}-R^{(1)}_{\mu\nu}],
\quad R^{(1)}=R^{(1)}_{\lambda}{}^{\lambda}=h_{,\lambda}{}^{\lambda}
-h^{\mu\nu}{}_{,\mu\nu},\quad h=h_{\lambda}{}^{\lambda}. \eqno(6)
$$
From (1-2) and (5-6) we find
$$
h=2\phi^2-4\phi-4+\frac2{1-\phi}+\frac2{1+\phi},\quad h_{,\lambda}{}^{\lambda}=
\frac4{r^2}\phi^2[\frac1{(1-\phi)^3}+\frac1{(1+\phi)^3}+1],
$$
$$
h_{ij}=(1-\phi)^2\delta_{ij}+\frac{\phi^2-\phi^3}{1+\phi}\frac{x_ix_j}{r^2}-
\delta_{ij},   \eqno(7)
$$
$$
h_{ij,kl}=\frac{2x_ix_jx_kx_l}{r^6}\left(-12\phi^2+15\phi-8+\frac3{1+\phi}
+\frac3{(1+\phi)^2}+\frac2{(1+\phi)^3}\right)+
$$
$$
\frac{2\delta_{ij}x_kx_l}{r^4}(4\phi^2-3\phi)+\frac{2\delta_{ij}\delta_{kl}}
{r^2}(\phi-\phi^2)+\frac{\delta_{ik}\delta_{jl}+\delta_{il}\delta_{jk}}
{r^2}\left(-\phi^2+2\phi-2+\frac2{1+\phi}\right)
$$
$$
 +\frac2{r^4}[x_ix_j\delta_{kl}+x_kx_j\delta_{il}+x_ix_k\delta_{jl}+
x_jx_l\delta_{ik}+x_ix_l\delta_{jk}]\times
$$
$$
\times\left(2\phi^2-3\phi+2-
\frac1{1+\phi}-\frac1{(1+\phi)^2}\right). \eqno(8)
$$
With the help of this expressions we obtain
$$
 R^{(1)}_{kl}=\frac{x_kx_l}{r^4}\left(2\phi^2-2+\frac1{1+\phi}+
  \frac1{(1+\phi)^2}-\frac1{1-\phi}-\frac1{(1-\phi)^2}+\frac2{(1-\phi)^3}
  \right)+
$$
$$
  \frac{\delta_{kl}}{r^2}\left(2-\frac3{1+\phi}+\frac1{(1+\phi)^2}
  +\frac1{1-\phi}-\frac1{(1-\phi)^2}\right),
$$
$$
R^{(1)}=\frac1{r^2}\left(2\phi^2+4+\frac4{1-\phi}-\frac8{(1-\phi)^2}+
\frac4{(1-\phi)^3}-\frac8{1+\phi}+\frac4{(1+\phi)^2}\right),\!\!\eqno(9)
$$
$$
t_{kl}=\frac1{8\pi G}[\frac{x_kx_l}{r^4}\left(-2\phi^2+2-\frac1{1+\phi}
-\frac1{(1+\phi)^2}+\frac1{1-\phi}+\frac1{(1-\phi)^2}
-\frac2{(1-\phi)^3}\right)+$$
$$
\frac{\delta_{kl}}{r^2}\left(\phi^2-\frac1{1+\phi}+\frac1{(1+\phi)^2}+
\frac1{1-\phi}-\frac3{(1-\phi)^2}+\frac2{(1-\phi)^3}\right)].\eqno(10)$$
For $\phi\ll1$ we have
$$
t_{ik}\vert_{\phi\ll1}=\frac{\phi^2}{8\pi
G}[\frac{7\delta_{ik}}{r^2}-\frac{14x_ix_k}{r^4}].\eqno(11)
$$
Similarly, we find
$$
t_{00}=\frac1{8\pi Gr^2}\left(\frac4{1+\phi}-\frac2{(1+\phi)^2}-\phi^2-
2\right), \eqno(12)
$$
$$
t_{00}\vert_{\phi\ll1}=-\frac3{8\pi G}(\nabla\phi)^2=-\frac{3GM^2}{8\pi r^4}.
\eqno(13)
$$
Now we check that conservation laws
 $t^{\mu\nu}{}_{,\nu}=0$ are fulfilled. As $t_{i0}=0$, we need to verify
 that $t_{ni,n}=0$.  Straightforward calculation gives
 $$
 R^{(1)}_{ni,n}=\frac12R^{(1)}_{,i}= $$
 $$
\frac{x_i}{r^4}[-4-4\phi^2+\frac4{1+\phi}+\frac4{(1+\phi)^2}-
\frac4{(1+\phi)^3}-\frac2{1-\phi}-\frac2{(1-\phi)^2}+\frac{10}{(1-\phi)^3}-
\frac6{(1-\phi)^4}],
$$
Q.E.D.

Introducing tensor
$$
Q^{\rho\nu\lambda}=\frac12[h^{,\nu}\eta^{\rho\lambda}-
h^{,\rho}\eta^{\nu\lambda}-h^{\mu\nu}{}_{,\mu}\eta^{\rho\lambda}+
h^{\mu\rho}{}_{,\mu}\eta^{\nu\lambda}+h^{\nu\lambda,\rho}-
h^{\rho\lambda,\nu}],\eqno(14)
$$
with property $Q^{\rho\nu\lambda}=-Q^{\nu\rho\lambda}$,
we have, see. Ch.7, \S6 ¢ [4]:
 $$
Q^{\rho\nu\lambda}{}_{,\rho}=R^{(1)\nu\lambda}-\frac12\eta^{\nu\lambda}R^{(1)}=
-8\pi G\tau^{\nu\lambda},
\eqno(15)
$$
$$
\tau^{\nu\lambda}=\eta^{\mu\nu}\eta^{\lambda\kappa}[T_{\mu\kappa}+t_{\mu\kappa}].
$$
 Due to this relation for smooth tensor
  $Q^{\rho\nu\lambda}$ the integral of total (gravitational and material)
  energy density over the volume of system  may be written as an integral over
  remote surface
  (see [4])
   $$
  P^0=-\frac1{8\pi G}\int Q^{i00}{}_{,i}d^3x=-\frac1{8\pi G}\int
  Q^{i00}n_ir^2d\Omega=M,\eqno(16)
 $$
$$
Q^{i00}=\frac12(h_{jj,i}-h_{ij,j})=\frac{x_i}{r^2}\left(2-\phi^2-\frac2{1+\phi}
\right),
$$
   $n_i$ are components of external normal to the surface.
Yet in case of horizon we see from (10) and (12) that at $\phi=-1$ (i.e. at
 $r=GM$ in harmonic system) tensor
$t_{\mu\kappa}$ has unintegrable singularity.
This prevent us from using Gauss theorem (i.e. from going to the second
equation in (16)) in the  whole volume of system. But it is easy to find the
energy  outside a sphere of radius $r_1=GM(1+\delta)$:
$$
\Delta P^0=-\left .\frac
r{2G}\left(2-\phi^2-\frac2{1+\phi}\right)\right |^{\infty}_{r_1} \eqno(17)
$$
For $0<\delta\ll1$ we get
$$
\Delta P^0=-M\left(\frac1{\delta}+\frac12\right).\eqno(19)
$$
It is interesting to compare (19) with Dehnen result [7]. Using standard \\
Schwarzschild coordinates ($r'=r+GM$) Dehnen find for his tensor
 $$
 \Delta
P^0=-GM^2\int_{r'_1}^{\infty}\frac{dr'}{r'^2\left(1-
\frac{r_g}{r'}\right)^{\frac32}}=M\left(1-
\frac1{\sqrt{1-\frac{r_g}{r'_1}}}\right).\eqno(20)
$$
This expression has square root divergence for $r'_1\to r_g$.

It is interesting that in Box 23.1 in [1] arguments are given  in favor
of localazability of gravitational energy density in case  of spherical
symmetry. According to this arguments the gravitational energy outside
the matter ball is zero. In Newtonian limit this way of accountig
for gravitational energy corresponds to accounting for gravitational attractions
between different parts of matter.  It is not clear how to reconcile this with
(19) or (20) and with the concept that nonlinear corrections to gravitational
field are generated by gravitational energy-momentum tensor. As to the
nonuniqueness of energy-momentum tensors, one may note that if some tensor
correctly describes the interaction with gravitons then it is natural to
consider this tensor as the right one.

Although Schwarzschild singularity is considered fictitious (from the time of
 Lema\^itre, see box 31.1 in [1]), it is difficult to be reconciled with this.
Physically the singularity manifest itself in impossibility for cosmonaut,
crossing it, to return back, in unlimited growth of acceleration (in static
frame) of freely falling particle nearing horizon, in absence of static frame
beyond horizon and in many other unusual things.

 One can consider these singularities as a hint that in the
regime of strong field the theory will be modified in the future and no
drastic changes in space-time topology will be needed.

The author is grateful to V.I.Ritus for useful discussions and constructive
   remarks.
\section{‹³×¥Á ×ÇÁ }
    \noindent 1.  Misner C.W., Thorne K.S., Wheeler J.A.,{\sl
Gravitation}, San Francisco (1973).  \\ 2.  Goldberg J.N., Phys.  Rev. {\bf111},
315 (1958).  \\ 3.  Fock V., {\sl The of Theory of Space-Time and Gravitation}
(2nd revised edition, Pergamon Press, New York, 1964).\\ 4.  Weinberg S., {\sl
Gravitation and Cosmology}, New York (1972).\\ 5. Novikov I.D., Frolov V.P.
     {\sl Physics of black holes,} Nauka, Moscow (1986).\\ 6.  A.I.Nikishov,
    gr-qc/9912034.\\ 7.  Dehnen H.  {Zeitschr.
   f\"ur Phys.}, {\bf179}, 76 (1964).  \end{document}